\documentclass[floatfix,aps,prf,onecolumn,showpacs,notitlepage,superscriptaddress,longbibliography]{revtex4-2}
\usepackage{graphicx}
\usepackage{xcolor}
\usepackage{amssymb}
\usepackage{ulem}
\usepackage{subcaption}

\begin{document}

\title{Drag reduction during the side-by-side motion of a pair of intruders in a granular medium}

\author{D. D. Carvalho}
\affiliation{Universit\'e Paris-Saclay, CNRS, Laboratoire FAST, 91405 Orsay, France}
\affiliation{Faculdade de Engenharia Mec\^anica, Universidade Estadual de Campinas (UNICAMP), Rua Mendeleyev, 200 Campinas, SP, Brazil}

\author{Y. Bertho}
\affiliation{Universit\'e Paris-Saclay, CNRS, Laboratoire FAST, 91405 Orsay, France}

\author{A. Seguin}
\affiliation{Universit\'e Paris-Saclay, CNRS, Laboratoire FAST, 91405 Orsay, France}

\author{E. M. Franklin}
\affiliation{Faculdade de Engenharia Mec\^anica, Universidade Estadual de Campinas (UNICAMP), Rua Mendeleyev, 200 Campinas, SP, Brazil}

\author{B. Darbois Texier}
\affiliation{Universit\'e Paris-Saclay, CNRS, Laboratoire FAST, 91405 Orsay, France}

\begin{abstract}
When several intruders move in a granular medium, coupling effects are observed, the motion of one intruder affecting that of others. In this paper, we investigate experimentally how the drag forces acting on a pair of spherical intruders moving amid grains at constant velocity vary with the transverse separation between them and their depth. When intruders are sufficiently far apart, they do not influence each other, and the average drag felt by each of them matches that of a single intruder. However, for small distances between intruders and at a given depth, the average drag per intruder decreases, highlighting a collaborative effect that facilitates motion. This collaboration effect is amplified when the depth of the intruders increases. We propose a model for the drag reduction of a pair of intruders based on the breakup of contact chains, caused by the perturbation generated by the neighbor intruder. Our findings provide new insights into the interaction effects on the motion of solids in sand, such as those observed in animal locomotion, root growth, and soil survey.
\end{abstract}

\maketitle

\section{Introduction}

The relative motion between solid bodies (intruders) and granular materials is ubiquitous in nature and human activities. In particular, at relatively small velocities, grains' motion is dominated by solid-solid friction and the formation and breakup of contact chains \cite{Kolb1, Tordesillas, Kozlowski, Carlevaro, Kozlowski2, Kozlowski3, Pugnaloni, Carvalho}, in a regime called quasistatic. This regime can be observed in the thrusting of plows in agricultural activities, in avalanche protection devices aimed at slowing down the flow \cite{Benito2012, DarboisTexier2023}, in root growth \cite{kolb2017physical, fakih2019root}, and in the motion of animals in the soil \cite{hosoi2015beneath}. One case of particular interest is when several intruders move within grains, since the motion of each intruder can affect those of others \cite{Carvalho2}, establishing a cooperative behavior. Further applications can be envisaged: if one intruder affects the motion of others, the ground can be probed to detect the presence of solid objects, such as buried rocks or ice. This opens new opportunities for prospecting the soil of planets and moons for the presence of ice and other materials, for instance.

In order to quantify the interaction between intruders moving through grains, various model experiments were carried out. The first type of study concerns intruders that move freely in low-density grains, corresponding to densities much lower than those of the intruders. For example, a pair of intruders impacting a light granular medium side by side first repel themselves in the horizontal plane at a low depth and attract each other at a higher depth \cite{Pacheco,Solano}. Whereas a Bernoulli-like mechanism is invoked to explain the attractive behavior, the repulsion is interpreted as being due to granular jamming in the region between the intruders. In a two-dimensional (2D) case, the behavior of several intruders was also studied numerically for pairs and trios of larger disks moving freely amid smaller disks \cite{Carvalho2}. These simulations showed the existence of cooperative dynamics between the intruders, even when they were at relatively large distances from each other. They also revealed that the final arrangement of the intruders in space depends on their initial positions, with particular cases where the same final configuration was reached for different initial conditions. The cooperative dynamics were rationalized as the result of compaction and expansion of granular matter in front and behind each intruder, respectively.

A second type of study concerns threaded objects placed at a constant separation distance, for which forces have been measured during their motion. For a pair of intruders, it has been shown that the side force acting on intruders varies with separation, from repulsive for small values to attractive for relatively higher values \cite{caballero2021attraction,de2016lift,dhiman2020origin}. Dhiman \textit{et al.} proposed a mechanism to explain these observations, where repulsion and attraction are given, respectively, by the formation and breakup of contact chains linking the intruders \cite{dhiman2020origin}. These chains depend, in turn, on the intruders' surfaces and the shear zones close to them: the intruders' surfaces tend to stabilize contact chains, while the shear zones tend to destabilize them. Moreover, Caballero \textit{et al.} found that the drag force acting on each intruder does not vary significantly with their separation, although it is lower than in the case of a single intruder \cite{caballero2021attraction}. For a bounded-2D granular system, Carvalho and Franklin \cite{Carvalho2} observed a nonmonotonic behavior of the drag force with the separation of the intruders. The drag force exhibits a minimum     at intermediate separation distance before returning to the same value as an individual intruder for a large separation.

Interaction effects between threaded objects were also observed for a pair of horizontal rods penetrating downward into a granular bed \cite{Pravin}. Beginning with contact intruders, the total mechanical work required for the penetration process first increases with the separation distance until the gap is large enough to allow the grains to flow between the rods. Once the grains can flow between the intruders, the total work decreases by 25\% until it reaches a plateau value. In order to characterize the interaction between two intruders, Merceron \textit{et al.} \cite{Merceron} carried out experiments where they visualized the granular flow around a pair of intruders placed side by side. The intruders are forced to move upward into a 2D granular medium consisting of small bidisperse disks. They showed that there is a separation distance between the intruders below which the motion of grains in front of one intruder is affected by the other. Remarkably, this distance is observed to be independent of the intruders' size.

Despite the progress made in recent years, many questions remain unanswered concerning the drag force resulting from these cooperative behaviors. Thus, we have conducted experiments in which we forced either one or two spheres to move horizontally at different depths within a granular medium while measuring the drag forces involved in their motion. We probe the spatial extent of the interaction between the spheres by exploring the influence of the separation distance on the mean drag experienced by each sphere and determining an associated characteristic length of the interaction. We also explore the possibility of developing a model to predict this drag.

After describing the experimental setup in Sec.~\ref{sec:setup}, we present in Sec.~\ref{sec:Res} drag force measurements of one or a pair of intruders moving horizontally through a granular medium. Section~\ref{sec:disc} presents a model to describe the reduction in drag as the distance between the intruders decreases and for different burial depths. Finally, Sec.~\ref{sec:conc} summarizes the main conclusions of this work.

\section{\label{sec:setup} Experimental setup}

\begin{figure}
  \centering
  \begin{subfigure}{0.54\textwidth}
    \includegraphics[width=\linewidth]{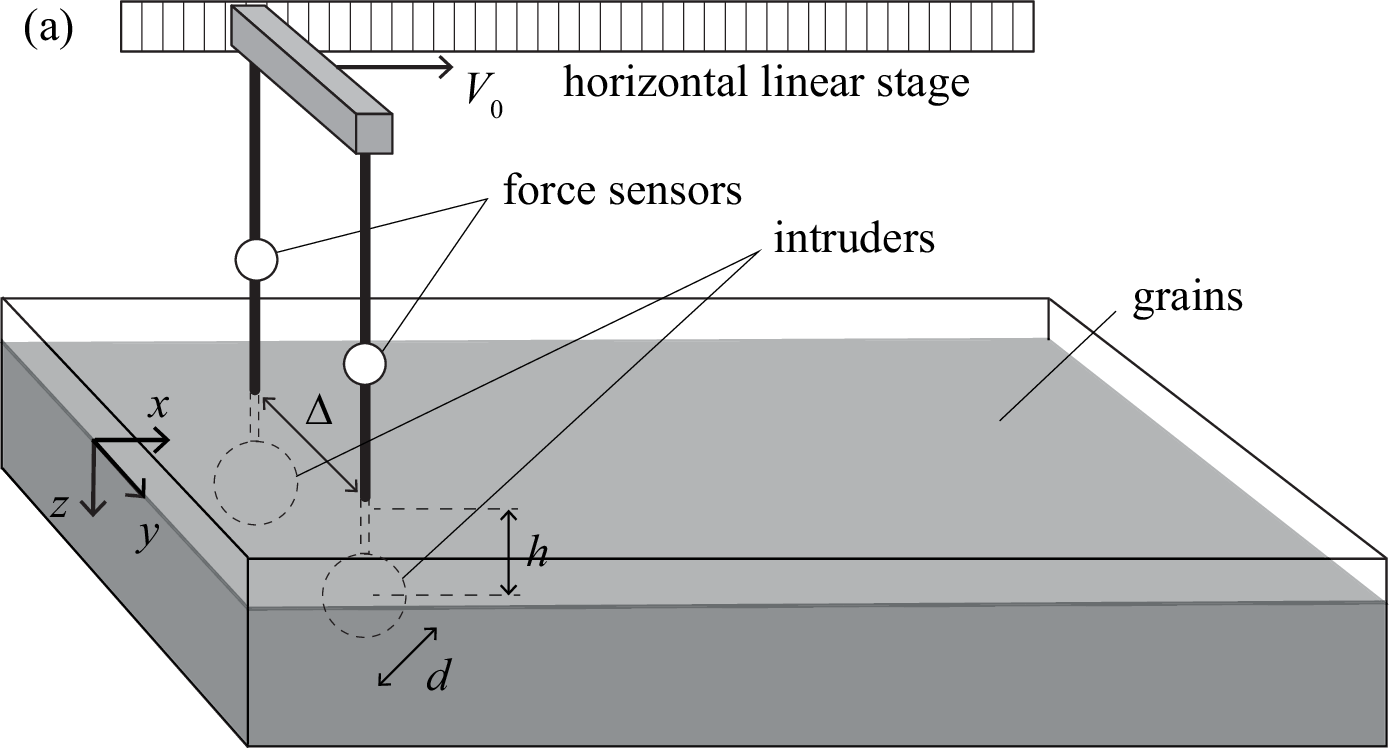}
  \end{subfigure}
  \hfill
  \begin{subfigure}{0.42\textwidth}
    \includegraphics[width=\linewidth]{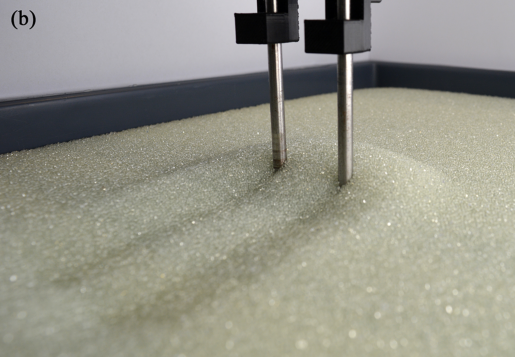}
  \end{subfigure}
  \caption{(a)~Sketch of the experimental setup for the displacement along the $x$ axis of two spherical intruders of diameter $d$, immersed in grains at the depth $h$. (b)~Picture of an experiment for two immersed intruders $\Delta=30$~mm apart, at depth $h=14$~mm and moving at the velocity $V_0=2.7$~mm~s$^{-1}$ during the forward journey. The image is used for visualization purposes to show the surface deformation that occurs at the shallowest depth.}
  \label{Fig01}
\end{figure}

The experiments consist of pulling intruders (polyamide spheres with a diameter $d=20$~mm) at a constant velocity $V_0$ inside a granular medium made of slightly polydisperse glass spheres (diameter $d_g = 1\pm 0.3$~mm and density $\rho\simeq 2.5\times 10^3$~kg~m$^{-3}$). The grains are contained in a rectangular box 365~mm long and 270~mm wide, and filled to a height of 97~mm [Fig.~\ref{Fig01}(a)]. To ensure the randomness of the initial conditions and the homogenization of the granular bed, the box is vibrated manually along the transverse $y$ direction before each experiment takes place, which results in a flattened free surface bed with an initial packing fraction of $\phi = 0.60 \pm 0.02$, measured by weighing the contents of the box. We find that this procedure leads to reproducible measurements. The intruders are attached to cylindrical rods of 5~mm in diameter, preventing any tilting or rotation, and immersed in the bed at depth $h$ ($h$ being the distance separating the free surface of grains from the center of the intruder). The rods are connected to force gauges (\textit{load cell 780~g, Phidgets Inc.}) that measure the longitudinal time-varying drag force $f_0(t)$ at a frequency of 60~Hz. The whole system is fixed to an $x$-direction moving plate, controlled by a linear stepper motor ensuring the displacement of the intruders at a velocity $V_0$ from 10$^{-1}$ to 10~mm~s$^{-1}$ [Fig.~\ref{Fig01}(a)].

Two distinct configurations will be considered below: (i)~the displacement of a single intruder in the $x$ direction from one edge of the box to the other, and initially placed at $y=0$ and depth $z=h$ ; (ii)~the displacement of two side-by-side intruders at the same abscissa $x$ and same depth $z=h$, located initially at $y = \pm \Delta/2$ and separated by a distance $\Delta$ measured from their centers as seen in Fig.~\ref{Fig01}(a). To prevent any wall effects \cite{Seguin2008}, we ensure to stay far enough from the side walls during an experiment, with a minimal intruder/wall distance of approximately $3\,d$, and we restrict immersion depths to $h\leq 49$~mm to maintain a distance greater than $2.5\,d$ between the intruders and the bottom wall. In addition, in the $x$ direction, data are acquired in a region of interest (ROI) located at the center of the box, in the range of 135~mm $\leq$ $x$ $\leq$ 225~mm. Finally, note that one experiment consists of moving the intruder forward (along positive $x$) in the undisturbed granular medium, and then, in a second step, making the return path (toward negative $x$) in the wake generated by the forward path and seen in Fig.~\ref{Fig01}(b). Note that free surface deformation can appear at shallow depths. The surface deformations become less significant as the object is deeper in the granular medium, as observed in previous studies \cite{gravish2010force}.

\section{\label{sec:Res} Experimental results}
\subsection{Journey of a single intruder}

\begin{figure}[t]
\centering
\includegraphics[width=\linewidth]{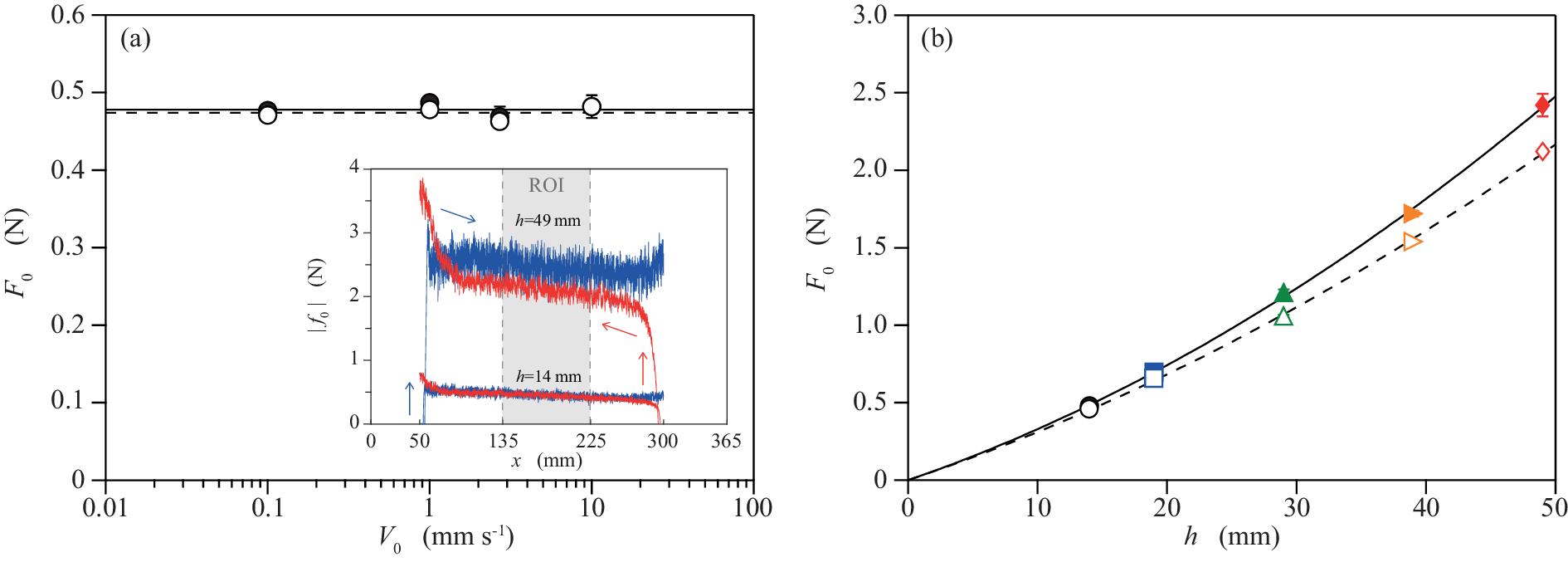}
\caption{(a)~Mean drag force on a single intruder $F_0$ as a function of its horizontal velocity $V_0$ at a depth $h=14$~mm. Solid symbols correspond to forward motion and open symbols to backward motion. Inset: Instantaneous drag force on a single intruder $f_0$ as a function of the position $x$ at two different depths ($h=14$~mm and $h=49$~mm), and a travel velocity of $V_0=2.7$~mm~s$^{-1}$ for (\textcolor{blue}{\rule{4mm}{0.5mm}})~ forward and (\textcolor{red}{\rule{4mm}{0.5mm}})~backward motions, respectively. The shaded area corresponds to the region of interest where the measurements are carried out. (b)~Mean drag force on a single intruder $F_0$ as it moves horizontally at $V_0=2.7$~mm~s$^{-1}$, as a function of immersion depth $h$. Solid symbols correspond to forward motion and open symbols to backward motion. The curves are the best fits of the data, of the form $F_0=A_0\,h+B_0\,h^2$, where the solid line corresponds to $A_0\simeq 2.9~10^{-2} \pm 0.1~10^{-2}$~N~mm$^{-1}$ and $B_0\simeq 4.1~10^{-4} \pm 0.3~10^{-4}$~N~mm$^{-4}$, and the dashed line to $A_0\simeq 2.8~ 10^{-2} \pm 0.1~10^{-2}$~N~mm$^{-1}$ and $B_0\simeq 3.1~10^{-4} \pm 0.2~10^{-4}$~N~mm$^{-2}$.}
\label{Fig02}
\end{figure}

\indent We first study the displacement of a single intruder at constant velocity $V_0$ in the granular medium at depth $h$. The inset in Fig.~\ref{Fig02}(a) shows the typical evolution of the intruder drag force $f_0$ as it moves at $V_0= 2.7$~mm~s$^{-1}$ at the depth $h=14$~mm, from one side of the box to the other. Note that here and in the remainder of this paper, the drag force measured during the movement of the rod alone (i.e., without the intruder attached to its end) has been subtracted from the force signal for each probed depth $h$, so as to retain only the force experienced by the intruder. This method has already been adopted in other experimental studies \cite{Albert2001}. Despite some fluctuations, associated with the creation/breaking of force chains \cite{Kolb1,seguin2016local,Carvalho}, we observe distinct zones in the force signal presented in the inset of Fig.~\ref{Fig02}(a). When the movement of the intruder begins, the force increases abruptly when the motion starts, and then, after a transient regime, reaches an extended zone of slow variation which becomes more negligible as the intruder is deeper in the granular medium \cite{seguin2019hysteresis}. Finally, on approaching the box walls, the force starts to increase again, as already reported in previous studies \cite{Kolb1,Carvalho}. In the following, we will define the mean drag force $F_0$ felt by a single intruder, as the average of the instantaneous force $f_0$ over the 90~mm long region of interest (ROI) located at the center of the box, so that $F_0=|\langle f_0(x)\rangle_\mathrm{ROI}|$, where the absolute value accounts for the mean drag force in both journeys of the intruders (forward and backward).

\indent Figure~\ref{Fig02}(a) displays the evolution of the mean drag force $F_0$ as a function of the imposed displacement velocity $V_0$. Drag force measurements are made during the intruder's first pass (solid symbols), and also during its backward return as it passes through the wake it previously created (open symbols). No significant variation in $F_0$ is observed with the displacement velocity $V_0$, over the two-decade range experimentally explored. In a granular medium, the fact that the drag force does not depend on velocity is the signature of a quasistatic regime characterized by a Froude number smaller than one. This Froude number, expressed as the ratio of the kinetic pressure due to collisions between grains to the pressure generated by the gravity field, is written as $\mathrm{Fr}=V_0/\sqrt{gh}$ and has a value smaller than 0.03 for all cases studied here, consistent with the quasistatic regime hypothesis \cite{Hilton2013,Takehara2010,Takehara2014,Faug2015,seguin2019hysteresis,Seguin2022}.

\indent Figure~\ref{Fig02}(b) shows the evolution of the mean drag force $F_0$ on a single intruder as a function of burial depth $h$. The drag force $F_0$ is observed to increase with the depth $h$, both in the forward and backward directions. This increase is slightly supralinear and can be modeled by a quadratic function of $h$, in agreement with observations from a previous experimental study \cite{Albert2001}. We also note that the backward drag force seems to be slightly lower than the forward drag force. This difference can be explained by the fact that the intruder passes in its wake, i.e., in an area that has been slightly structured in some way by its first passage \cite{guillard2013depth}. Note that the horizontal free surface has also been disturbed by the passage of the rod, slightly modifying the effective burial height.

\subsection{Journey of two side-by-side intruders}

\begin{figure}[t]
\centering
\includegraphics[width=0.5\linewidth]{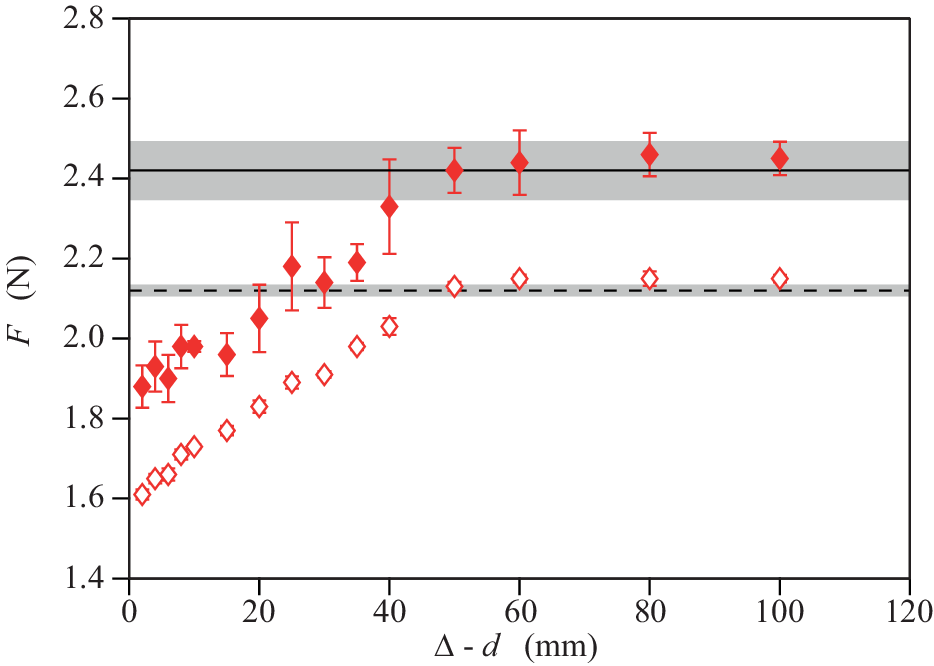}
\caption{Mean drag force $F$ for both intruders as a function of the space between them $\Delta -d$, for a displacement at $V_0=2.7$~mm~$^{-1}$ at the depth $h=49$~mm. Solid symbols correspond to forward motion and open symbols to backward motion, and error bars represent the standard deviation in five realizations. The horizontal lines and shaded areas correspond, respectively, to force values $F_0$ and their typical fluctuations, obtained with a single intruder under similar conditions.}
\label{Fig03}
\end{figure}

Let us now consider the displacement of two side-by-side intruders separated by a distance $\Delta$, at velocity $V_0$ and depth $h$, as sketched in Fig.~\ref{Fig01}(a). Each force sensor provides a signal similar to that shown in the inset in Fig.~\ref{Fig02}(a). The overall drag force $F$ of the system composed of the two intruders is determined as the average of the mean drag forces experienced by each of the intruders, $F_0$. Figure~\ref{Fig03} shows the evolution of the mean drag force $F$ as a function of the distance between the intruders ($\Delta-d$) while the spheres velocity and depth were kept constant to $V_0=2.7$~mm~$^{-1}$ and $h=49$~mm respectively. Each point in this figure corresponds to an average of five experiments. The error bars correspond to the standard deviation calculated on these five realizations. The variation of force $F$ with distance between intruders ($\Delta-d$) follows the same trend, for both the forward and backward journeys. For large separations ($\Delta-d \gtrsim 50$~mm), the mean drag force $F$ remains constant. This constant value corresponds to that measured for a single intruder, on both the forward and backward journeys, as attested by the horizontal lines in Fig.~\ref{Fig03} depicting the drag force $F_0$ for a single intruder under similar conditions. Thus, the two intruders do not interact with each other. When the intruders are closer to each other ($\Delta-d \lesssim 50$~mm), the average drag force is significantly lower than at large separation. This decrease can reach up to 30\% in relative value when the distance between intruders vanishes, which is well below the usual force fluctuations observed for a single intruder (depicted as gray regions in Fig.~\ref{Fig03}). Therefore, they cooperate with each other, resulting in a reduction in drag force. Finally, as already mentioned for a single intruder, we note that the drag force is about 10\% lower during the backward journey than during the forward one due to the passage of intruders in their own wake. We also observe that the error bars are smaller in the case of backward motion than in the case of forward motion.

\begin{figure}[t]
\centering
\includegraphics[width=\linewidth]{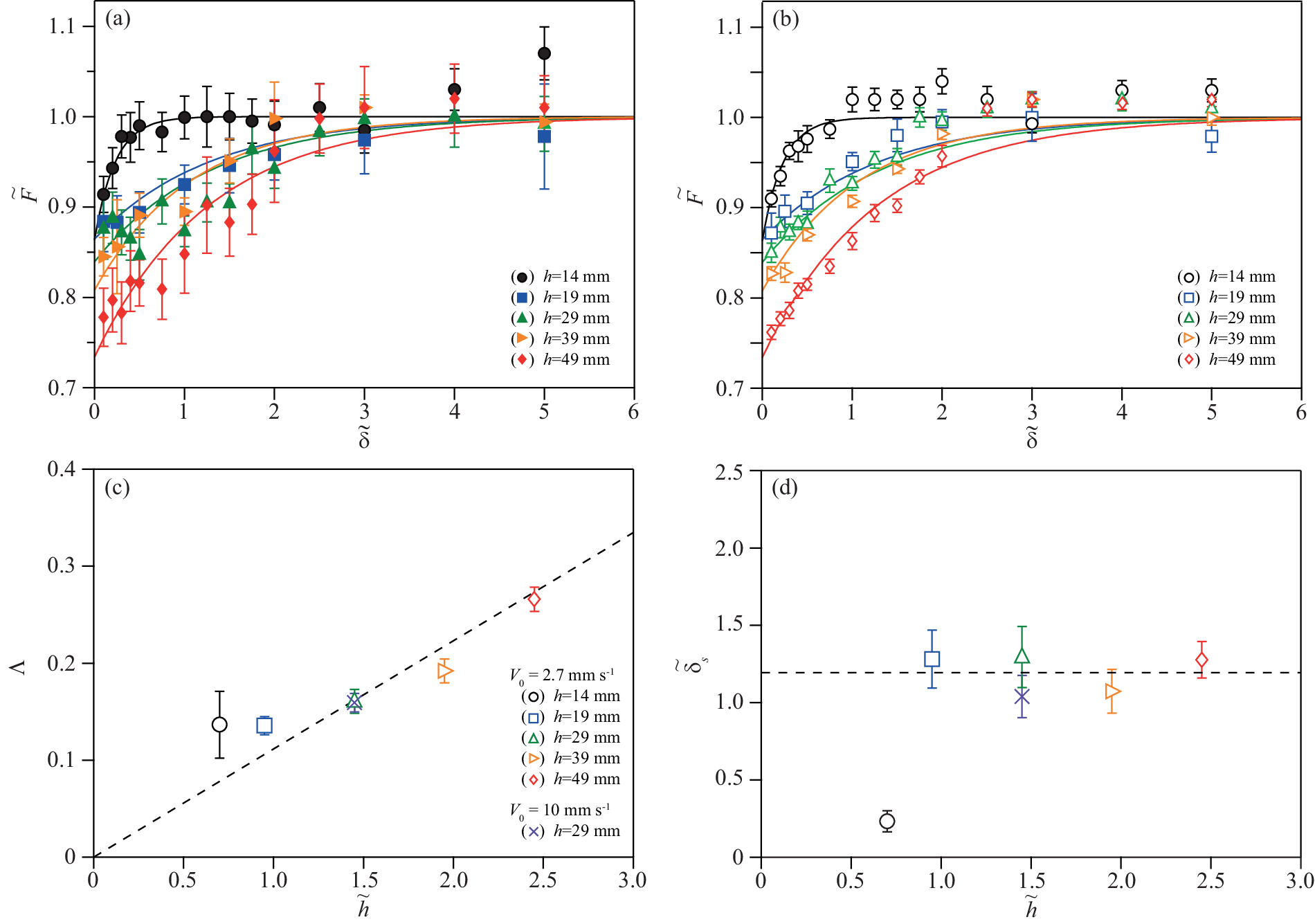}
\caption{Normalized drag force $\tilde{F}=F/F_0$ as a function of the normalized distance between the intruders $\tilde{\delta}=(\Delta-d)/d$ for (a)~a forward motion and (b)~a backward motion at $V_0=2.7$~mm~s$^{-1}$, and different depths ($\bullet$,$\circ$)~$h=14$~mm, (\textcolor{blue}{$\blacksquare$}, \textcolor{blue}{$\square$})~$h=19$~mm, (\textcolor{green}{$\blacktriangle$}, \textcolor{green}{$\vartriangle$})~$h=29$~mm, (\textcolor{orange}{$\blacktriangleright$}, \textcolor{orange}{$\triangleright$})~$h=39$~mm, (\textcolor{red}{$\blacklozenge$}, \textcolor{red}{$\lozenge$})~$h=49$~mm. Solid lines correspond to the best fits of the data with Eq.~(\ref{eq:exp}). Parameters (c)~$\Lambda$ and (d)~$\tilde{\delta}_s$ resulting from the fitting of the data with Eq.~(\ref{eq:exp}) as a function of the normalized depth $\tilde{h}$. Dashed lines correspond to (c)~$\Lambda=0.1\tilde{h}$ and (d)~$\tilde{\delta}_s=1.2$. In the panels, error bars represent the standard deviation in five realizations.}
\label{Fig04}
\end{figure}

Figure~\ref{Fig04} shows the normalized drag force $\tilde{F}=F/F_0$ as a function of the normalized distance between intruders $\tilde{\delta}= (\Delta -d)/d$ for different burial depths $h$, on both the forward [Fig.~\ref{Fig04}(a)] and backward [Fig.~\ref{Fig04}(b)] journeys. The evolution of $\tilde{F}$ with $\tilde{\delta}$ is similar in both cases and also whatever the burial depth $h$. It can also be seen that when the two intruders are close enough to each other, the greater the burial depth, the greater the relative drag reduction. Since the force increases roughly linearly with $\tilde{\delta}$ at short distances between intruders, and saturates at a constant value when they are far enough apart, we propose to model the observed behavior with an exponential law of the form
\begin{equation}
\tilde{F}=1-\Lambda\exp\left(-\frac{\tilde{\delta}}{\tilde{\delta}_s}\right),
\label{eq:exp}
\end{equation}
where $\Lambda$ is a coefficient corresponding to the relative reduction in drag and $\tilde{\delta}_s$ a normalized characteristic screening length reflecting the typical distance from which intruders can affect each other. When $\tilde{\delta}<\tilde{\delta}_s$, the intruders cooperate and the drag force per intruder is lower than the value for a single intruder moving through the grains. Conversely, when $\tilde{\delta}>\tilde{\delta}_s$, the intruders do not interact with each other, i.e., we recover the case of a single intruder. Note that even in the limit of touching intruders, the separation distance between the two rods holding the intruders is large enough to ensure that there is no interaction between them.

Equation~(\ref{eq:exp}) allows us to fit our experimental data for each burial depth $h$ shown in Figs.~\ref{Fig04}(a) and \ref{Fig04}(b), and extract the corresponding $\Lambda$ and $\tilde{\delta}_s$ values. We note that the solid lines plotted in Figs.~\ref{Fig04}(a) and \ref{Fig04}(b) were obtained for both the forward and backward journeys together. The evolution of the drag reduction coefficient $\Lambda$ and the normalized screening length $\tilde{\delta}_s$ as a function of the normalized burial depth $\tilde{h}=h/d$ are plotted in Figs.~\ref{Fig04}(c) and \ref{Fig04}(d), respectively. It can be seen that as the penetration depth increases, the $\Lambda$ reduction is greater. For the sake of simplicity, the whole data set can be described by a linear behavior of the form $\Lambda \sim \tilde{h} $. The linear fit proposed here should not be valid for larger values of $\tilde{h}$, as it is unreasonable to expect $\Lambda$ to exceed the value 1. Saturation of $\Lambda$ is therefore expected for large values of $\tilde{h}$ which are above the depth range possible to be explored with the current experimental setup. The normalized screening length $\tilde{\delta}_s$ is observed to be rather constant with the normalized burial depth $\tilde{h}$, with $\tilde{\delta}_s=1.2$. We also note that for the shallowest depth ($\tilde{h}\simeq 0.7$), the normalized screening length $\tilde{\delta}_s$ and reduction $\Lambda$ deviate from these trends. These deviations may be due to a free-surface effect since they are observed when the burial depth is less than one sphere diameter. It is important to note in Figs.~\ref{Fig04}(c) and \ref{Fig04}(d) that additional measurements were carried out at a velocity approximately four times higher (cross symbols in these figures), and we observe that both the coefficients $\Lambda$ and $\tilde{\delta}_s$ are unaffected, at least in the quasistatic regime.

\begin{figure}[t]
\centering
\includegraphics[width=0.5\linewidth]{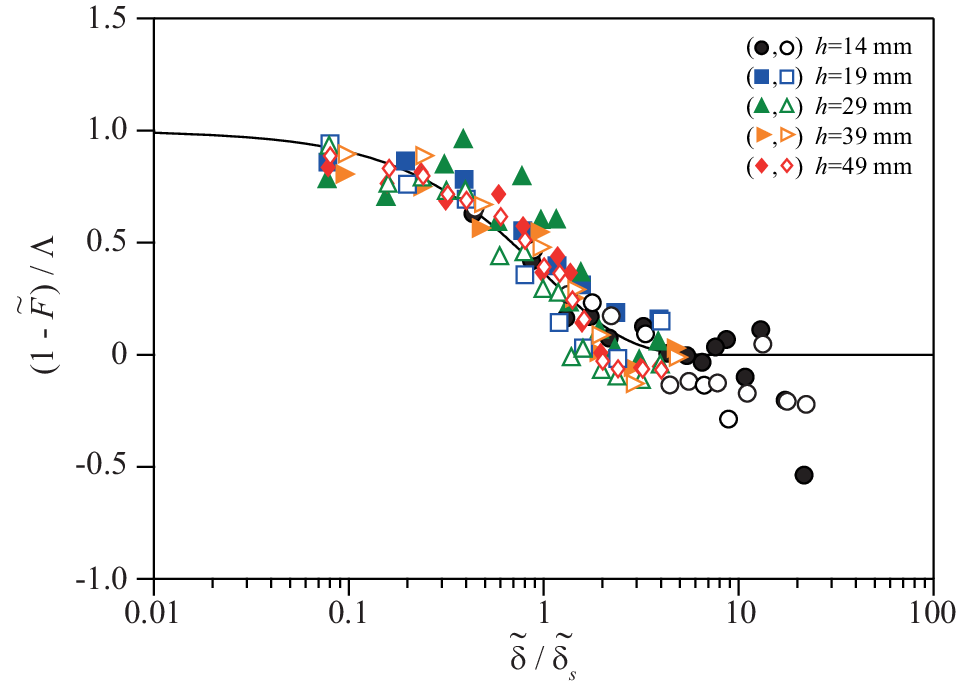}
\caption{Rescaled drag force $(1 - \tilde{F}) / \Lambda$ as a function of the rescaled separation distance $\tilde{\delta}/\tilde{\delta}_s$. For each set of data, we use the values of $\Lambda$ and $\tilde{\delta}_s$ resulting from the best fits obtained in Fig.~\ref{Fig04}. All the data collapse on the solid line which corresponds to Eq.~(\ref{eq:exp}).}
\label{Fig05}
\end{figure}

Finally, it is possible to propose a master curve on which the data are superposed. Figure~\ref{Fig05} shows the evolution of $(\tilde{F}-1)/\Lambda$ as a function of $\tilde{\delta}/\tilde{\delta}_s$. We can see that the data gather around the experimental fit of Eq.~(\ref{eq:exp}) for both the forward and backward runs. We also observe that measurements at shallow depths deviate from the model due to free-surface effects.

\section{\label{sec:disc} Discussion}

\begin{figure}[t]
\centering
\includegraphics[width=\linewidth]{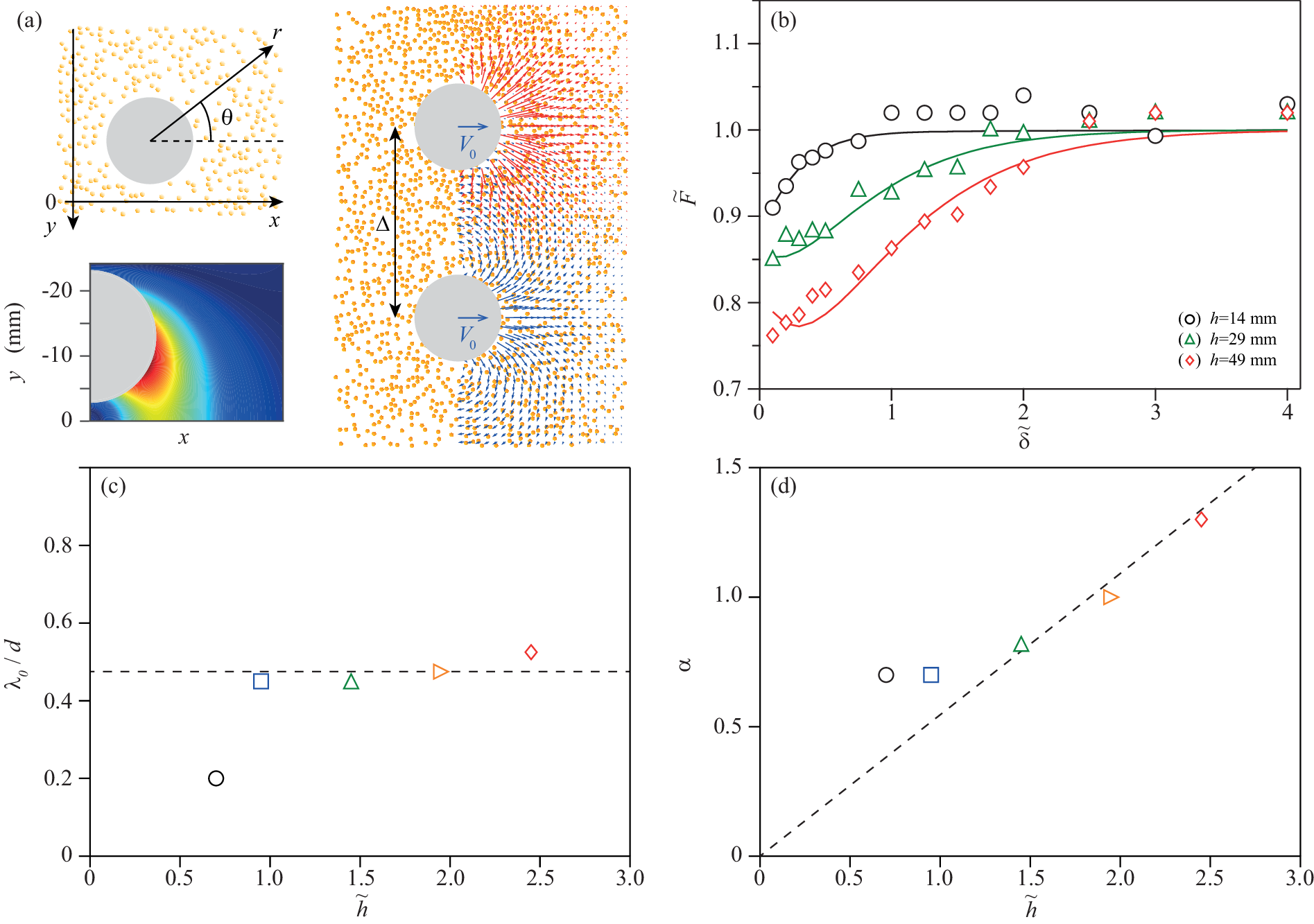}
\caption{(a)~Top view sketches illustrating the model developed to rationalize the interaction between the two intruders: (top left)~notations of the problem; (bottom left)~area in front of one intruder where the color scale from blue to red encodes increasing values of the interaction term $\alpha|\mathbf{v}_1 \cdot \mathbf{v}_2|/V_0 ^2$ in Eq.~(\ref{eq:force_modified}); (right)~velocity field (blue arrows) of the neighboring intruder at the bottom disturbs the force field (red arrows) induced by the movement of the intruder at the top (and vice versa, not shown here for visibility purposes). (b)~Normalized drag force $\tilde{F}$ for a backward motion as a function of the normalized separation distance $\tilde{\delta}$. Symbols correspond to experimental data at different depths and solid lines represent the best fit with the model. (c)~Parameter $\lambda_0$ divided by $d$ resulting from the best fit of the data as a function of the normalized depth $\tilde{h}$. The dashed line indicates $\lambda_0/d\simeq 0.48$. (d)~Coefficient $\alpha$ resulting from the best fit of the data as a function of the normalized depth $\tilde{h}$. The dashed line corresponds to $\alpha\simeq 0.55\,\tilde{h}$.}
\label{Fig06}
\end{figure}

In this section, we discuss how the interactions between two intruders moving side by side in a granular material can be rationalized. For the side force appearing on two side-by-side cylinders in a granular flow, it has been shown experimentally and numerically that the direction of this force (attraction or repulsion) correlates with the sign of the difference in granular temperature between the inside and outside of the cylinders \cite{de2016lift, caballero2021attraction}. The same argument of granular temperature difference has been invoked to rationalize the axial segregation of large spheres in a rotating drum filled with small grains \cite{zuriguel2005role}. However, Dhiman \textit{et al.} \cite{dhiman2020origin} carried out discrete element method (DEM) simulations of two side-by-side intruders and showed that the temperature and pressure fields do not follow the evolution expected by kinetic theory. They concluded that the difference in granular temperature should be a consequence of the interactions rather than its cause. In order to understand the origin of the side forces that appear on the intruders when they are close to each other, Dhiman \textit{et al.} \cite{dhiman2020origin} studied numerically the dynamics of force chains in their vicinity. They found that the presence of a neighbor shears the force chains of the first intruder and breaks them more often. As a result, the first intruder pushes less on the granular material on its neighbor's side than it does on the other side. In addition, this would also predict a reduction of the drag force on the intruder when its neighbor is close. This scenario is also in agreement with the observations of Reddy \textit{et al.}, who observed that the presence of a shear zone in the vicinity of a cylindrical intruder immersed in grains reduces its yielding force \cite{reddy2011evidence}. We therefore formulate the observations of Dhiman \textit{et al.} \cite{dhiman2020origin} in an empirical model where the force field of the first intruder is perturbed by the velocity field of its neighbor. Our approach is two-dimensional and is based on time-averaged local force and velocity fields within the granular material, which is described as a continuous medium. The velocity field around a cylinder moving amid grains has been studied experimentally \cite{seguin2013experimental} with the following relation:
\begin{equation}
\frac{\mathbf{v} (r,\theta)}{V_0} = -A_r (r) \cos \theta \, \mathbf{e}_r + A_\theta(r) \sin \theta \, \mathbf{e}_\theta,
\label{eq:velocity_field}
\end{equation}
where $r$ and $\theta$ are the cylindrical coordinates with origin at the center of the object, as shown in Fig.~\ref{Fig06}(a) and the two functions $A_r$ and $A_\theta$ write:
\begin{equation}
A_r (r) = \frac{r - d/2 + \lambda_s}{r} \left [1 - \exp \left({- \frac{r - d/2}{\lambda_0}} \right)\right ] \quad \mathrm{and} \quad A_\theta (r) = 1 + \frac{r - d/2 + \lambda_s - \lambda_0}{\lambda_0} \exp \left({- \frac{r-d/2}{\lambda_0}}\right),
\label{eq:velocity_field2}
\end{equation}
where $\lambda_0$ is the characteristic length over which the velocity varies along the radial direction, and $\lambda_s$ reflects the velocity slip tangential to the object surface. These two parameters have been shown to depend on the cylinder diameter $d$ and the grain size $d_g$ according to the empirical relations: $\lambda_0 = d/4+ 2d_g$ and $\lambda_s = 0.45 \, d$  \cite{seguin2013experimental}. In our case, i.e., an intruder of 20~mm in diameter moving amid grains of 1~mm, it corresponds to $\lambda_0/d = 0.35$ and $\lambda_s/d = 0.45$. Note that the velocity field given by Eq.~(\ref{eq:velocity_field}) is expressed in the reference frame of the intruder, and here it should be expressed in the reference frame of the laboratory by adding $+ V_0 \, \mathbf{e}_x$ to Eq.~(\ref{eq:velocity_field}). This velocity field is represented with blue arrows in Fig.~\ref{Fig06}(a). The network of forces around an object moving in a granular medium has been studied in experiments with photoelastic grains \cite{Clark2012} or by numerical simulations \cite{Carvalho}. Both approaches reveal that the force distribution extends over a characteristic length: high stresses are applied close to the object and there are no stress variations far from the intruder \cite{Muthuswamy2006}. Furthermore, experiments with 2D photoelastic grains revealed that the force distribution decreases exponentially with the distance from the intruder \cite{Clark2012,seguin2016local}. From these studies, it is possible to propose an expression for the average local force per unit area in the granular material, which follows an exponential decay and whose pattern resembles that of the force chains usually observed around intruders \cite{seguin2016local,Carvalho}. The force per unit area around the intruder can be described empirically by the following expression:
\begin{equation}
\mathbf{f} =f_r \exp\left(-\frac{r-d/2}{\lambda_0}\right)\cos\theta\,\mathbf{e}_r+ f_\theta \exp\left(-\frac{r-d/2}{\lambda_0}\right) \sin\theta\,\mathbf{e}_{\theta}.
\label{eq:force_field}
\end{equation}
This expression assumes that the characteristic length over which the force field varies in the radial direction is the same as that of the velocity field, i.e., $\lambda_0$. It also introduces two force coefficients, $f_r$ and $f_\theta$, whose ratio reflects the way the force chains deviate from the radial direction. The average force field of Eq.~(\ref{eq:force_field}) corresponds to the space average of the normal forces transmitted by contacts between particles per unit area and is a continuous representation of the discrete contact network. Figure~\ref{Fig06}(a) shows with red arrows a typical example of a force field resulting from Eq.~(\ref{eq:force_field}). In this approach, the total drag force experienced by an intruder is considered to correspond to the integral of the force field per unit area over a surface bounded by the perimeter of the intruder and a thickness of one grain. We have only considered the frontal part of the object because it is the main contributor to the total drag force \cite{seguin2016local}. Thus the drag force $F_0$ expresses as
\begin{equation}
    F_0 = \int_{-\pi/2} ^{\pi/2}\int_{d/2} ^{d/2+d_g}\, {\mathbf{f} \cdot \mathbf{e}_x}\,\,dr\,rd\theta,
    \label{eq:drag_force}
\end{equation}
and can be calculated analytically as
\begin{equation}
    F_0 = \frac{\pi}{2} (f_r- f_\theta) \lambda_0 \left[\lambda_0+\frac{d}{2}-\left(\lambda_0+\frac{d}{2} + d_g \right) \exp\left(-\frac{d_g}{\lambda_0}\right) \right].
    \label{eq:drag_force_exp}
\end{equation}
Note that in the limit where $d_g \ll \lambda_0$, the previous expression simplifies to $F_0 \simeq (\pi/4)(f_r-f_\theta)dd_g$. Under these circumstances, the drag force scales linearly with the effective surface area of the object $d d_g$, as expected in two-dimensional configurations \cite{seguin2019hysteresis, Seguin2022}.

In the rest of the paper, the forces will be normalized by this reference value $F_0$ given by Eq.~(\ref{eq:drag_force_exp}), which corresponds to the case where there is no interaction, and we will consider the ratio $\tilde{F}=F/F_0$ introduced in Sec.~\ref{sec:Res}. In order to account for the mechanism proposed by Dhiman \textit{et al.} \cite{dhiman2020origin} in this framework and calculate the drag force in the presence of interactions, we assume that the force field around the first intruder $\mathbf{f}_1$ is perturbed locally by the velocity field of the second intruder $\mathbf{v}_2$. We hypothesize that the perturbed force field $\mathbf{f}_1'$ writes
\begin{equation}
    \mathbf{f}_1 ' = \mathbf{f}_1 \left(1 - \alpha \frac{|\mathbf{v}_1 \cdot \mathbf{v}_2|} {V_0 ^2}\right),
    \label{eq:force_modified}
\end{equation}
where $\alpha$ is a nondimensional coefficient that represents the strength of the interaction. In this phenomenological formulation, $\alpha$ is an ad hoc parameter adjusted from experimental data. Note that this expression depends only on the orientation of the local velocity fields of the two intruders and not on their norm. The area of velocity and force interactions between the two intruders is highlighted in Fig.~\ref{Fig06}(a).
In the following, we study how this interaction modifies the total drag force on one intruder as a function of the separation distance $\Delta$. We solve this problem numerically by computing Eq.~(\ref{eq:drag_force}), where the force field $\mathbf{f}$ is replaced by the perturbed force field $\mathbf{f}_1'$ given by Eq.~(\ref{eq:force_modified}). This calculation gives the drag force $F$ in the presence of interaction and allows it to be compared with the reference force $F_0$. We repeat this procedure for different separating distances $\Delta$ and compute the resulting drag force ratio $\tilde{F}$ on one intruder. This approach results in a normalized drag force $\tilde{F}$ that tends toward one for large separation distances ($\Delta \gg d$) and decreases as the separation distance is reduced, in agreement with our observations. As we have used renormalization, the drag force ratio $\tilde{F}$ is only a function of $\alpha$, $\lambda_0$ and $\lambda_s$. Therefore, we consider $\lambda_0$ and $\alpha$ as free parameters, keep $\lambda_s/d=0.45$ constant, and search for the best fits of the measurements of the normalized drag force at each depth in the case of a backward motion. Figure~\ref{Fig06}(b) presents the normalized drag force as a function of the separation distance for three different depths along with the best fits of the model. We observe that the model correctly captures the reduction in drag force observed experimentally. The estimated values of $\lambda_0/d$ and $\alpha$ found with this procedure are plotted as a function of $\tilde{h}$ in Figs.~\ref{Fig06}(c) and \ref{Fig06}(d), respectively. Note that similar parameters are found when adjusting the forces measured in the case of forward motion. We observe in Fig.~\ref{Fig06}(c) that the ratio $\lambda_0/d$ is roughly independent of the depth of the intruder, except at shallow depths where surface effects are not negligible. At large depths, the ratio $\lambda_0 /d$ is about 0.48, in close agreement with the estimate of Seguin \textit{et al.} \cite{seguin2013experimental}. Figure~\ref{Fig06}(d) shows that the coefficient $\alpha$ resulting from the fitting procedure increases linearly with depth. This observation is consistent with the fact that the parameter $\alpha$ is related to the relative magnitude of drag reduction, which follows the same evolution as seen in Fig.~\ref{Fig04}(c).

In addition, this procedure can also be used to calculate the side force experienced by the intruders when they interact with each other. The side force is calculated by integrating the force field projected onto the $y$ direction in a similar way to Eq.~(\ref{eq:drag_force}) as: $F_S = \int_{-\pi/2} ^{\pi/2} \int_{d/2} ^{d/2+ d_g} \, {\mathbf{f}_1 ' \cdot \mathbf{e}_y}\,dr\,rd\theta$. In doing so, we find that the lateral force tends toward zero at large separation distances ($\Delta \gg d$) and is attractive at smaller separation distances. This prediction is in line with previous observations made in experiments and simulations for separation distances that are not too small ($\Delta > 0.1 d$) \cite{de2016lift, dhiman2020origin, caballero2021attraction}. Since side forces are not the scope of this work, we decided to not present these results here. Finally, the framework proposed above, although empirical, provides an extensive prediction of the interaction between two objects moving in a granular material in the quasistatic regime.

Another point that is relevant to discuss is the reference case of two spheres in interaction in a viscous fluid, and how the situation compares with the granular case. The interaction in viscous fluids has been analytically solved using force-point methods by Happel and Brenner \cite{happel1983low}. In the case of two identical spheres moving side by side at a constant velocity (flow perpendicular to the centerline linking the spheres), the drag force is predicted to reduce as the separation between the spheres decreases. At the first order, the drag force on the sphere is predicted to reduce as $F / 3 \pi \eta d V_0 = 1 /[ 1+(3/8) d/\Delta ]$ where $\eta$ is the fluid's viscosity. Note that in the limit of touching spheres ($\Delta = d$), the relative drag reduction in the viscous case is approximately equal to 27\%, which is the same order of the highest drag reduction measured in our experiments in granular materials [Fig.~\ref{Fig04}(c)]. Such drag reduction in viscous flow has been observed experimentally in the case of two bubbles ascending side by side \cite{saad2015interactions}, and with two spheres in yield-stress fluid \cite{merkak2006spheres}. However, in the case of interacting spheres in viscous flows, no depth dependence on drag reduction is predicted. Another important difference between both situations is that side forces are not present in the viscous case.

\section{\label{sec:conc}Conclusion}
In this paper, we investigated experimentally how drag forces acting on a pair of transversely aligned intruders vary with their depth and transverse separation as they move at constant speed in a granular bed. We found that the mean drag experienced by each intruder is lower than that for a single intruder when separations are small, and that the drag increases with their separation, until it reaches a plateau equal to the single intruder's value for large separations, evincing, therefore, a cooperation dynamics within a given distance range. In addition, we found that the drag reduction for small separations increases with depth and that data for the mean drag varies exponentially with the intruder-intruder separation, and propose a model for the drag reduction based on the breakup of contact chains caused by the local motion of grains. Despite the progress made so far, some aspects need to be investigated further, such as the forces on intruders when they are closer than one grain diameter of each other, since small distances were limited in our experimental setup, or are at higher depths than those presented in this paper, given that one should expect a saturation in drag reduction at large depths. Although the model has some limitations, being two-dimensional and using phenomenological laws, it describes well our experimental results. Another perspective would be to study the influence of free surface deformations at shallow depths. Other examples are the behavior of intruders when their motion is not confined to the longitudinal direction, so that they can approach or repel each other, move upward, or even rotate, and the system behavior for higher speeds and different shapes of intruders and grains. Our findings, however, shed light on the cooperative dynamics and coupling effects taking place in granular media.

\begin{acknowledgments}
The authors thank J.~Amarni, A.~Aubertin, L.~Auffray, C.~Manquest and R.~Pidoux for their technical support. This work has benefited from fruitful discussions with G. Gauthier, H. Perrin, M. Rabaud and F. Melo. The authors are grateful to the S\~ao Paulo Research Foundation FAPESP (Grants No. 2018/14981-7, No. 2020/04151-7, and No. 2022/12511-9) for financial support.
\end{acknowledgments}

\bibliography{biblio}

\end{document}